\documentclass[]{raa}            
\usepackage{graphicx,times}
\usepackage{natbib}

\providecommand{\bm}[1]{\mbox{\boldmath$#1$\unboldmath}}

\begin{document}

   \title{A Critique of Supernova Data Analysis in Cosmology
}

 \volnopage{ {\bf 2010} Vol.\ {\bf 10} No. {\bf 12}, 000--000}
   \setcounter{page}{1}

   \author{Ram Gopal Vishwakarma
      \inst{1}
   \and Jayant V. Narlikar
      \inst{2}
   }

   \institute{Unidad Acad$\acute{e}$mica de Matem$\acute{a}$ticas,
 Universidad Aut$\acute{o}$noma de Zacatecas,
 C.P. 98068, Zacatecas, ZAC,
 Mexico; {rvishwa@mate.reduaz.mx}\\
        \and
             Inter-University Centre for Astronomy and Astrophysics, Pune
411007, India; {jvn@iucaa.ernet.in}\\
\vs \no
   {\small Received [year] [month] [day]; accepted [year] [month] [day] }
}

\abstract{ Observational astronomy has shown significant growth over the last decade and has made important contributions to cosmology. A major paradigm shift in cosmology was brought about by observations of Type Ia supernovae. The notion that the universe is accelerating has led to several theoretical challenges. Unfortunately, although high quality supernovae data-sets  are being produced, their statistical analysis leaves much to be desired. Instead of using the data to directly  test the model, several studies seem to concentrate on assuming the model to be correct and limiting themselves to estimating model parameters and internal errors. As shown here, the important purpose of testing a cosmological theory is thereby vitiated. 
\keywords{cosmology: observations - supernovae Ia: general
}
}

   \authorrunning{R. G. Vishwakarma \& J. V. Narlikar }            
   \titlerunning{A Critique of Supernova Data Analysis in Cosmology }  
   \maketitle


%
%

\noindent
Observations play a crucial role in all branches of science. Moreover, they are 
more important in cosmology where, on one hand, the events are non-repeatable 
and, on the other hand, the theoretical side is more speculative than laboratory physics, requiring guidelines from the observations by 
confronting them. The power of observations in cosmology is clear from the 
observations of supernovae (SNe) of type Ia, which dramatically changed, about 
a decade ago, the then standard picture of cosmology -  of an expanding universe evolving under the rules of general relativity such that the expansion rate
should slow down as cosmic time unfolds. Amazingly, the
first generation of data showed that the rate of expansion of the universe is
speeding up! This gave rise to a plethora of models to explain the agent driving the acceleration, as well as some possible modifications of general relativity. 
Theorists may debate the relative merits of various cosmic-acceleration
theories: cosmological constant, dark energy, alternative gravity, anthropic 
arguments, etc., but it is ultimately up to the observations to decide which theory is correct.

However, we notice a recent unfortunate trend in the analysis of SNe Ia data, which departs from an objective assessment of a theory by observations. Everybody would agree that the first step in fitting the observational data to a theory is to check whether the theory is consistent with the data for viable values of its free parameters (if any). In order to do this, there are two standard ways: (i) the Bayesian approach, which gives a relative rather than an absolute measure of how good a theory is, and hence is 
 more appropriate for comparison between competing theories; (ii) the maximum likelihood approach, which is more commonly used for hypothesis testing.  Under this approach, one minimises $\chi^2$ which measures the deviations of the theoretical predictions from the observations. In the present case of the observations
of magnitude ($m$) and redshift ($z$) of SNe Ia, the $\chi^2$ is  given by the value
\be
\chi^2 = \sum_{j = 1}^N \,\frac{[m_{\rm t}(z_j;~{\rm parameters}) - m_{{\rm o}, j}]^2}
{\sigma_{m_j}^2},\label{eq:chi}
\ee
where $m_{\rm t}(z_j;~{\rm parameters})$ is the theoretical value of the magnitude at redshift $z_j$ of the $j$th supernova predicted by a model which is given in terms of its parameters.  The observed magnitude of the $j$-th SN  is $m_{{\rm o}, j}$. The variance $\sigma_{m_j}^2$ represents the combined uncertainty in the observed magnitude of the $j$th supernova arising from the uncertainties in the different variables, for example, lensing, dust extinction, peculiar velocity of the host galaxy, etc. By Taylor expanding $m$ around its mean value and by recalling that the variance of $(m)\equiv \sigma_m^2=\langle m^2\rangle-\langle m\rangle^2$, one can write the combined uncertainty of $m$ in terms of the uncertainties in its variables, say, $x_i$:
\be
\sigma_m^2=\sum_{i}\left(\frac{\partial m}{\partial x_i}\right)^2\sigma_{x_i}^2+\sum_{i}\sum_{k\neq i}\left(\frac{\partial m}{\partial x_i}\right)\left(\frac{\partial m}{\partial x_k}\right){\rm cov}(x_i,x_k),\label{eq:errors}
\ee
where ${\rm cov}(x_i,x_k)$ is the covariance between the variables $x_i$ and $x_k$, which vanishes for any uncorrelated variables.
It is obvious from equation (\ref{eq:chi}) that if the model  satisfactorily represents
the data, the difference between the predicted magnitude and 
the observed one at each data point should be roughly the same size as the 
measurement uncertainties and each data point would contribute roughly one to $\chi^2$, giving the sum that roughly equals
the number of data points $N$ (more correctly $N-$number of fitted parameters
$\equiv$ number of degrees of freedom `DoF').
If $\chi^2$ is much larger, the fit is considered bad. A more quantitative assessment of the goodness-of-fit is given in terms of the $\chi^2$-probability.  
If the fitted model
provides a typical value of $\chi^2$ as $x$ at $n$ DoF, this probability is
given by
\be
P(x, n)=\frac{1}{\Gamma (n/2)}\int_{x/2}^\infty e^{-u}u^{n/2-1} {\rm d}u.
\ee
Thus, $P(x, n)$ gives the probability that a model which does fit the data at $n$ DoF, would give a value of $\chi^2$ as large
as $x$ or larger. Qualitatively, $P$ represents the probability of finding a worse fit to the data.
If $P$ is very small, the fit is not acceptable.
For example, if we get a $\chi^2=20$ at 4 DoF for some model, then the 
hypothesis that {\it the model describes the data satisfactorily} is  unlikely, 
as the probability $P(20, 4)=0.0005$ is very small.
However, the $\chi^2$-probability $P$ holds strictly only when the measurement errors are normally distributed. In reality, they are not. In most cases, the effect of non-Gaussian errors is to create an abundance of outlier points, which decrease the probability $P$. It is due to this reason that
usually models with $P>0.001$ are considered acceptable. 
One should proceed to estimate the parameters of the model only after examining if the model has a credible goodness-of-fit, in the absence of 
which, the estimated parameters of the model (and their estimated 
uncertainties) have no meaning
at all.

Nevertheless, we noticed (see also Vishwakarma 2007) that recently an approach has been followed to analyze the SNe Ia data which does not respect the standard procedure described above. Initiated by Astier et al. (2006), this approach simply assumes (rather than examines) that the standard cosmology (23\% of dark matter and 72\% of dark energy) is consistent with the SNe Ia observations and limits itself to parameter estimation. Under this new approach, $\chi^2$ is calculated from 
\be
\chi^2 = \sum_{j = 1}^N \,\left[ \frac{\{m_{\rm t}(z_j;~{\rm parameters}) - m_{{\rm o}, j}\}^2}
{\sigma_{m_j}^2+\sigma_{\rm int}^2}\right],\label{eq:chinew}
\ee
where $\sigma_{\rm int}$ is the (unknown) intrinsic dispersion of the 
SN absolute magnitude (sometimes termed $\sigma_{\rm sys}$, unknown systematic uncertainties) which is used as an adjustable free parameter 
in order to obtain $\chi^2$/DoF = 1.
Clearly this new approach does not test the model under consideration. Rather, it {\it assumes} that the model is correct and goes on to find the best value for intrinsic dispersion. In principle, any model applied to the data in this way will determine $\sigma_{\rm int}$ and the purpose of testing the model or estimating its parameters is lost.

One may consider $\chi^2$/DoF = 1 to estimate $\sigma_{\rm int}$ from the nearby SNe alone as $\sigma_{\rm int}$ estimated from the nearby SNe alone and that estimated from the whole sample, do not differ significantly. For example, $\sigma_{\rm int}=0.15 \pm 0.02$ estimated 
from the fit of the nearby sample ($z \leq 0.1$) of Astier et al. (2006) appears to be statistically 
consistent with $\sigma_{\rm int}=0.13 \pm 0.02$ estimated from their whole 
sample and $\sigma_{\rm int}=0.12 \pm 0.02$ estimated from their distant SNe 
alone. However, one can always argue why a cosmological theory should be 
assumed to be correct (i.e., $\chi^2$/DoF = 1) even for the nearby SNe, even though 
all the Robertson-Walker models reduce to the $m_{\rm t}=5\log z+\rm constant$ for
low z.
In addition, a more reasonable way to introduce $\sigma_{\rm int}$ in the theory, is to use independent measurement
uncertainties $\sigma_{\rm int, j}$ from the individual SNe and use the correct expression for $\sigma_m^2$ given by equation (\ref{eq:errors}) where $\sigma_{\rm int}$ enters through the term
$(\partial m/\partial x_{\rm int})^2 \sigma_{\rm int}^2$.

Of course, one can estimate $\sigma_{\rm int}$ (if one is just interested in that) from all (high- as well as low-redshift) SNe data by assuming that a particular theory (here the standard cosmology), already tested, must be consistent with the data (i.e., $\chi^2/{\rm DoF}=1$). This is fine if the theory is well established and is already
tested by other independent ways. Then we are not interested in testing the already established  theory, but rather we want to estimate, from(\ref{eq:chinew}), some parameters of the data (here $\sigma_{\rm int}$) which we could not decipher from the observations. This procedure is followed in many
branches of physics. But what are the other observations which predict dark energy
independently and conclusively? To date, it is only the SNe Ia observations (if taken independently, i.e., not combined with other observations). The only other precise observations are
the anisotropy measurements of the CMB made by the WMAP project. However,
taken at their face values, the only relevant prediction of the WMAP
observations is of a flat geometry, and the decelerating models (like the Einstein-de Sitter model: $\Omega_m=1$, $\Lambda=0$) also successfully explain them (see for example, Blanchard 2003; Blanchard et al. 2005; Vishwakarma 2003). Observations on gravitational lensing, quasars, galaxy
clusters, gamma ray bursts, etc., are not precise enough. 
Age considerations depend heavily on the observations of $H_0$
and $\Omega_m$ which have wide-ranging degeneracy. Moreover, the best estimates of these parameters also include those which are estimated from
the SNe Ia observations themselves. In fact, a conclusive prediction of dark energy only comes from the SNe Ia case. To make the situation worse, there have been claims (Wei 2010) of some conflicts between recent SNe Ia data sets Union (Kowalski et al. 2008) and Constitution (Hicken et al. 2009); these disagree not only with CMB and BAO (baryon acoustic oscillation) studies, but also with other SNe Ia data sets. 

So, if we do not have
conclusive independent evidence for dark energy, what is the meaning of
$\sigma_{\rm int}$ (and other parameters) estimated from this model, following this new approach to data analysis?
The current concordance model in cosmology (also known as $\Lambda$CDM cosmology) has 18 parameters
(one can extract these parameters from Hinshaw et al. 2009 and Kowalski et al. 2008),
17 of which are independent. Thirteen of these parameters are well fitted to
the observational data; the other four remain floating. This situation is
very far from healthy, not only because of very few observational tests which support the standard cosmology, that it calls for more observational tests, but also  because of  the controversial theoretical aspects.  The well known fine tuning and coincidence problems related to dark energy, as well as its extremely speculative character (particle physics even fails to identify a theory for dark energy), have  invoked possible modifications of general relativity. 
Additionally on the observational side, there remains a lack of understanding of some issues related to SNe Ia observations causing a number of systematic uncertainties (which are likely to depend on redshift) that could affect the use of SNe Ia as standard candles in such cosmological probes. For example, one can mention the evolution of luminosity in SNe Ia (Dominguez et al. 1998; Hoflich et al. 1998; Drell et al. 2000; Timmes, Brown \& Truran 2003)  
and extinction of SNe light by dust which is still a poorly understood phenomenon (Holwerda 2008; Albrecht et al. 2006; Conley et al. 2007). Finally, it may be noted that different methods for analyzing SNe Ia data, in order to estimate magnitudes,
do not seem entirely consistent with each other, causing systematic error concerns. For example,
two different light-curve fitters, the multi-color light-curve shape (MLCS) method and the spectral-adaptive light-curve template (SALT) method, give significantly different estimates of magnitudes
of the individual SNe; besides, the magnitudes estimated by the SALT fitter acquire a degree of
cosmology-dependence (Frieman 2008).
 This situation warrants more rigorous tests of the theories as well as the observations. Certainly it suggests caution in regard to accepting the standard model as proven. 

 Initiated by the SuperNova Legacy Survey ``SNLS''  (Astier et al. 2006) and followed by Union1 (Kowalski et al. 2008), Union2 (Amanullah et al. 2010), Sloan Digital Sky Survey-II  ``SDSS-II'' (Kessler et al. 2009) and SNLS 3-year (Guy et al. 2010) data sets, this flawed approach has already acquired the status of `standard practice' in Lampeitl et al. (2010). The harmful side effect of the current approach is that even if the authors do not follow this approach, usually they do not care to check if their model fits the data or not, and limit themselves to estimating the parameters of the model either by only using the SNe Ia data or by combining the SNe Ia data with other observations. This is 
clear from the results of the Constitution data (Hicken et al. 2009), the largest sample so far consisting of 397 SNe Ia: 
although the authors do not follow the new approach, they do not seem to notice that the theory does not fit the data well! 
 One can calculate from their Table 1 that the best-fitting $\Lambda$CDM model, with $\Omega_{\rm m}$=$1-\Omega_{\Lambda}$= 0.29,  gives $\chi^2/{\rm DoF}=465.5/395$ with a meager probability $P=0.83\%$; and so the estimated model can be ruled out at more than a 99\% confidence level! Other cosmologies also have a similar fit.

Finally, we emphasize that our arguments here offer a criticism of the statistical technique used for analyzing the data rather than a criticism of the standard model itself.




\bigskip
\noindent
{\bf References}
\medskip

\noindent
Albrecht A. et al., 2006, astro-ph/0609591.\\
\noindent
Amanullah R., et al. 2010, ApJ. 716, 712.\\
\noindent
Astier P., et al., 2006, A \& A, 447, 31.\\
\noindent
Blanchard A., et al., 2003, A \& A, 412, 35.\\
\noindent
Blanchard A., 2005, `Cosmological Interpretation from High Redshift Clusters Observed Within the 

XMM-Newton $\Omega$-Project', {\it Proceedings of DARK 2004, the Fifth International Heidelberg 

Conference, October 3-9, Texas A\&M University} [preprint: astro-ph/0502220].\\
\noindent  
Conley A., Carlberg R. G., Guy J., Howell D. A., Jha S., Riess A. G., Sullivan M., 2007, ApJ 664, L13.\\
\noindent
Dominguez I., et al, `Type Ia Supernovae: Influence of the 
   Progenitor on the Explosion', {\it Proceedings 

of ``Nuclei in the Cosmos V" 
   Volos, Greece} (1998) [preprint: astro-ph/9809292].\\
\noindent
Drell P. S., Loredo T. J. and Wasserman I., 2000, ApJ, 530, 593.\\
\noindent
Frieman, J. A. 2008, in AIP Conf. Ser. 1057, 87 (arXiv: 0904.1832).\\
\noindent
Guy, J., et al. 2010, arXiv: 1010.4743.\\
\noindent
Hicken M., et al., 2009, ApJ, 700, 1097.\\
\noindent
Hinshaw G., et al., 2009, ApJ. Suppl., 180, 225.\\
\noindent
Hoflich P., Wheeler J. C. \& Thielemann F. K., 1998, ApJ, 495, 617.\\
\noindent 
Holwerda B. W., 2008, MNRAS, 386, 475.\\
\noindent
Kessler R., et al., 2009 (to appear in ApJS), arXiv:0908.4274.\\
\noindent
Kowalski M., et al., 2008, ApJ, 686, 749.\\ 
\noindent
Lampeitl H., et al., 2009, (to appear in MNRAS), arXiv:0910.2193.\\
\noindent
Timmes F. X., Brown E. F. \& Truran J. W., 2003, ApJ, 590, L83.\\
\noindent
Vishwakarma R. G., 2003, MNRAS, 345, 545.\\
\noindent
Vishwakarma R. G., 2007, Int. J. Mod. Phys. D, 16, 1641.\\
\noindent
Wei H., 2010, Phys. Lett. B, 687, 286.\\


\end{document}